\documentclass[twocolumn,showpacs,preprintnumbers,amsmath,amssymb]{revtex4-1}

\usepackage{graphicx}
\usepackage{dcolumn}
\usepackage{bm,color}

\newcommand{\Tr}[1]{\underset{#1}{\mbox{Tr}}}
\newcommand{\Eref}[1]{eq.~(\ref{#1})}
\newcommand{\Fref}[1]{Fig.~\ref{#1}}
\newcommand{\figwidth}{3.35in}

\begin{document}

\preprint{APS/123-QED}

\title{Replica symmetry breaking in an adiabatic spin-glass model of adaptive evolution}

\author{Ayaka Sakata}
\email{ayaka@sp.dis.titech.ac.jp}
\altaffiliation[Present address: ]{Department of Computational
Intelligence and Systems Science, Tokyo Institute of Technology,
Midori-ku, Yokohama 226-8502, Japan.}
\affiliation{Department of Basic Science, Graduate School of Arts and
Sciences, The University of Tokyo, Komaba, Meguro-ku, Tokyo 153-8902, Japan.}
\author{Koji Hukushima}
\email{hukusima@phys.c.u-tokyo.ac.jp}
\affiliation{Department of Basic Science, Graduate School of Arts and
Sciences, The University of Tokyo, Komaba, Meguro-ku, Tokyo 153-8902, Japan.}
\author{Kunihiko Kaneko}
\email{kaneko@complex.c.u-tokyo.ac.jp}
\affiliation{Department of Basic Science, Graduate School of Arts and
Sciences, The University of Tokyo, Komaba, Meguro-ku, Tokyo 153-8902, Japan.}

\date{\today}

\begin{abstract}
We study evolutionary canalization
using a spin-glass model with replica theory, where spins and their
interactions are dynamic variables whose configurations correspond
to phenotypes and genotypes, respectively.
The spins are updated under temperature $T_S$,
and the genotypes evolve under temperature $T_J$, according to the evolutionary fitness.
It is found that adaptation occurs
at $T_S< T_S^{\rm RS}$,
and a replica symmetric phase emerges at $T_S^{\rm RSB} < T_S <
 T_S^{\rm RS}$.
The replica symmetric phase implies canalization,
and replica symmetry breaking at lower temperatures indicates loss of robustness.
%
\end{abstract}

\maketitle

%
Biological evolution occurs through changes in genotypes and phenotypes over generations,
driven by random genetic variance and natural selection.
This process preferentially selects genotypes that produce a phenotype that affords high
evolutionary fitness \cite{Futuyma,
Hartl-Clark}.
Thus, phenotypes, such as protein expression levels or the functional structures of proteins,
are the result of dynamic processes governed by the genes. However, such processes generally involve stochasticity due to
thermal noise, and thus phenotypes of isogenic individuals
are not necessarily identical\cite{Elowitz,Kaern,Furusawa}.
Indeed, such phenotypic fluctuations and the possible role of noise
have been extensively investigated both experimentally \cite{Sato,Laundry} and
theoretically \cite{Kaneko-PLoS, Wagner, SHK}.

For a phenotype to conserve its function, however,
it must be robust to this noise, at least to some degree.
Indeed, the dynamic adaptation process that shapes phenotypes exhibits global and
smooth attraction, as observed in the folding dynamics of proteins
\cite{Go, Onuchic}, RNA \cite{Ancel-Fontana},
protein expression dynamics 
governed
by gene regulatory networks \cite{Li},
developmental dynamics \cite{KK-Asashima},
and so forth. Besides this robustness to noise, the adapted phenotype
should be robust to genetic change to acquire evolutionary stability.
The possible relationship between these two types of robustness,
as well as the positive role of noise,
has recently been investigated theoretically\cite{Ancel-Fontana,Sasai-Yomo,Wagner,Kaneko-PLoS,SHK}.
The study found a transition toward robustness in the dynamic process with respect to the noise level (temperature), where the energy landscape for the dynamics changes from being rugged to having a funnel-like structure.

Considering the above change in the dynamical process,
one may expect that loss of robustness could be viewed as a transition
to the spin-glass phase in statistical physics.
Thus far, however, no analytic theory to support this view has been provided,
and, from a theoretical standpoint, little is understood of this transition in the evolution of robustness against noise (temperature).

Here we introduce a simple statistical-mechanics model
of adaptive evolution to explain the dynamical process that shapes phenotypes. We use
an adiabatic two-temperature spin-glass model
in which the spin configuration and the interaction matrix correspond
to the phenotype and genotype, respectively.
The genotype evolves to increase fitness
which is defined by the spin configuration.
With an analysis based on replica theory,
we demonstrate the emergence of a replica-symmetry-breaking transition
as the temperature decreases,
and show that the transition corresponds to a loss of robustness in the phenotype.
Adaptive evolution of robustness is shown to occur only in the replica symmetry phase,
where the Hamiltonian for global attraction to the adapted phenotype
is represented in terms of frustration. We also discuss the significance of replica symmetry breaking on phenotypic robustness.

Let us consider a simple spin model in which the
phenotype and genotype are represented by configurations of spin variables $\bm{S}\equiv\{S_i\}$
and the interaction matrix elements $\bm{J}\equiv\{J_{ij}\}$, respectively, with $i,j=1,\cdots,N$.
Each spin variable $S_i$ can take one of two values, $\pm 1$.
The interactions are fully connected between two spins.
Both the spins $\bm{S}$ and
interactions $\bm{J}$ are treated as dynamical variables, but the
time scale associated with $\bm{J}$ is much slower
so that  the interactions are relatively fixed during the time
evolution of the spins.
Thus, the equilibrium distribution of the spins is given by
$P(\bm{S}|\bm{J})=\frac{1}{Z_S(\bm{J})}\exp(-\beta_SH(\bm{S}|\bm{J}))$,
where $\beta_S=T_S^{-1}$, 
the Hamiltonian is given by 
$H(\bm{S}|\bm{J})=-\frac{1}{\sqrt{N}}\sum_{i<j}J_{ij}S_iS_j$,
and $Z_S(\bm{J})$ is a partition function under a given $\bm{J}$.
Within the long evolutionary time scale for $\bm{J}$,
the spin configuration driven by a
Hamiltonian $H_J$ reaches thermal equilibrium.
The distribution function of $\bm{J}$ is given by
$P(\bm{J})=\frac{1}{Z_J}\exp(-\beta_JH_J(\bm{J}))$,
where $\beta_J=T_J^{-1}$ 
and
$Z_J$ is the total partition function.
The function $H_J$ 
is generally expressed in terms of equilibrium
quantities of $\bm{S}$ and a bare distribution $P_0(\bm{J})$.
Here we set the Hamiltonian of $\bm{J}$ as
\begin{align}
H_J(\bm{J})=-\Psi(\bm{J})-T_J\log P_0(\bm{J}),
\end{align}
where $\Psi(\bm{J})$ is a fitness function.
The bare distribution is given by 
$P_0(\bm{J})=\prod_{i<j}\frac{1}{\sqrt{2\pi}}\exp\left(-J_{ij}^2\slash
(2J_0^2)\right)$,
with a unit $J_0$ of the interaction.
We assume that fitness is determined by a specific
configuration of given
$t$ spins, called target spins here.
(For example, protein function depends on the conformation of a set of residues,
and is indeed modeled by the configurations of target spins in \cite{Sasai-Yomo}).
More specificically, we assume that a functional phenotype is generated
when the configurations of target spins satisfy $\sum_i^tS_i=t\mu$
with $\mu$ being a constant value.
The remaining $N-t$ spins, called non-target spins, have no
direct influence on the selection of individuals.
The fitness function is thus defined by
\begin{align}
\Psi(\bm{J})=\log\Big\langle\delta\Big(\mu,\frac{1}{t}\sum_{i=1}^tS_i\Big)\Big\rangle\equiv\log\langle\psi(\bm{S})\rangle,
\label{fitness}
\end{align}
where
$\delta$ is Kronecker's delta and
$\langle\cdots\rangle$ is the thermal average with respect to the
spin variables according to the equilibrium distribution. 
The fitness function $\Psi(\bm{J})$ implies a logarithmic
probability for the magnetization of $t$-spins to take the value $\mu$
in equilibrium.
Note that it does not matter which spins are chosen as targets
because the model is a fully-connected
mean-field model.
The configuration of $t$-spins is not important either, because of the
gauge symmetry, which guarantees that a system with any
configuration of $t$-spins can be transformed into the system studied here,
without altering the thermodynamic properties \cite{Nishimori}.
The equilibrium distribution of $\bm{J}$ and
the total partition function are written as
\begin{align}
P(\bm{J})&=\frac{1}{Z_J}P_0(\bm{J})\langle\psi(\bm{S})\rangle^{\beta_J},~~~~Z_J=[\langle\psi\rangle^{\beta_J}]_0,
\label{prob_dist}
\end{align}
where $[\cdots]_0$ means the average over
$\bm{J}$ with respect to the bare distribution $P_0(\bm{J})$.
When $T_J=\infty$ or $t=0$,
the distribution $P(\bm{J})$ is identical to $P_0(\bm{J})$
irrespective of their fitness values,
and the system corresponds to the Sherrington--Kirkpatrick (SK) model.
For finite $T_J$ and $t$,
the interactions $\bm{J}$ that frequently lead to the spin configuration
with $\sum_{i=1}^t S_i=t\mu$
appear with higher probability.
In this sense, the temperature $T_J$ plays the role of the selection pressure
in genotypic evolution.


Assuming that $\beta_J$ is a positive integer,
the quantity $\langle\psi(\bm{S})\rangle^{\beta_J}$
can be expressed in terms of $\beta_J$ real replicas.
Following the replica method \cite{beyond},
the total partition function $Z_J$ can be expressed as
\begin{align}
Z_J=\lim_{n\to 0}\int
 D\bm{J}P_0(\bm{J})\Tr{\{\bm{S}^\alpha\}}\prod_{\alpha=1}^{\beta_J}\psi_\alpha e^{-\beta_S\sum_{\gamma=1}^nH_\gamma},
\label{Z_J-replica}
\end{align}
where $\psi_\alpha=\psi(\bm{S}^\alpha)$ 
and $H_\alpha=H(\bm{S}^\alpha|\bm{J})$. 
The right hand side of eq.~(\ref{Z_J-replica}) is originally calculated
for a positive integer $n$ and $\beta_J$
while keeping $\beta_J$ smaller than $n$,
and then the partition function $Z_J$ is analytically continued to non-integer $\beta_J$ and
non-integer $n$ with the limit to 0.
After some 
calculations,
the total free energy can be derived as a function
of replicated order parameters $\{q_{\alpha\beta}\}$,  their
conjugate parameters $\{\hat{q}_{\alpha\beta}\}$, and parameters
$\{\hat{\mu}_\alpha\}$ conjugated with $\mu$, which are determined by
self-consistent equations.
The replicas from the first to $\beta_J$-th are subjected to the external field $\psi_\alpha$,
and the others are not.
Taking the difference in the replicas into account,
we introduce a replica symmetric (RS) assumption for $q_{\alpha\beta}$ as
\begin{align}
q_{\alpha\beta}=\left\{
\begin{array}{ll}
q_1,~\mbox{if}&\alpha\leq\beta_J,\beta\leq\beta_J\\
q_2,~\mbox{if}&\alpha\leq\beta_J,\beta>\beta_J~\mbox{or}~\alpha>\beta_J,\beta\leq\beta_J\\
q_3,~\mbox{if}&\alpha>\beta_J,\beta>\beta_J.
\end{array}
\right.
\label{q_RS}
\end{align}
For the conjugate parameters $\{\hat{\mu}^\alpha\}$, it is assumed that
$\hat{\mu}_\alpha=\hat{\mu}$ for any $\alpha\leq\beta_J$.
With these assumptions, the RS total free energy density $f_{\rm RS}$
is given by
\begin{align}
\nonumber
f_{\rm RS}&(T_S,T_J,\mu)=-p\mu\beta_J\hat{\mu}-\frac{\beta_J\beta_S^2(q_1-q_3)}{2}\\
\nonumber
&-\frac{\beta_S^2}{2}\Big(\frac{\beta_J(\beta_J-1)q_1^2}{2}-\beta_J^2q_2^2
+\frac{\beta_J(\beta_J+1)q_3^2}{2}\Big)\\
&
+(1-p)\log \Xi(0)+p\log \Xi(\hat{\mu}),
\end{align}
where $p=t\slash N$.  
Here $\Xi(w)$ is defined as a normalization constant of the distribution
\begin{align}
P(u,v;w)&=\frac{e^{-(u^2+v^2)\slash
 2}}{2\pi\Xi(w)}\Big(\frac{\cosh(w+\sqrt{\hat{q}_1}u)}{\cosh
 W(u,v)}\Big)^{\beta_J},
\label{eqn:Iweight}
\end{align}
where
$W(u,v)=\sqrt{\frac{\hat{q}_2^2}{\hat{q}_1}}u+\sqrt{\frac{\hat{q}_1\hat{q}_3-\hat{q}_2^2}{\hat{q}_1}}v$;
and $\hat{q}_1$, $\hat{q}_2$, and $\hat{q}_3$ are the conjugate
parameters of $q_1$, $q_2$, and $q_3$, respectively.
At $\mu=0$, the free energy is identical to that of the SK
model under the RS ansatz.
The self-consistent equations for
the order parameters $q_1$, $q_2$, and $q_3$ are given by
\begin{align}
\nonumber
q_1&=(1-p)\langle\tanh^2(\sqrt{\hat{q}_1}u)\rangle_0\\
&\hspace{2.0cm}+p\langle\tanh^2(\hat{\mu}+\sqrt{\hat{q}_1}u)\rangle_{\hat{\mu}}\label{saddle_q1_2}\\
\nonumber
q_2&=(1-p)\langle\tanh(\sqrt{\hat{q}_1}u)\tanh
 W(u,v)\rangle_0\\
&\hspace{1.0cm}+p\langle\tanh(\hat{\mu}+\sqrt{\hat{q}_1}u)\tanh
 W(u,v)\rangle_{\hat{\mu}}\label{saddle_q2_2}\\
q_3&=(1-p)\langle\tanh^2W(u,v)\rangle_0+p\langle\tanh^2W(u,v)\rangle_{\hat{\mu}},\label{saddle_q3_2}
\end{align}
where
$\langle\cdots\rangle_x$ denotes the average according to the distribution
(\ref{eqn:Iweight}) at $w=x$.
The conjugate parameters of $q_i$s are given by
$\hat{q}_i=\beta_S^2q_i~(i=1,2,3)$.
The first and second terms of the order parameters come from
the non-target spins and the target spins, respectively.
Thus, eqs.(8)--(10) can be rewritten 
as the summation of the non-target and the target parts,
$q_i=(1-p)q_i^{\rm nt}+pq_i^{\rm t}~(i=1,2,3)$.
%
The
conjugate parameter $\hat{\mu}$ is implicitly determined by the equation
\begin{align}
\mu =
 \langle\tanh(\hat{\mu}+\sqrt{\hat{q}_1}u)\rangle_{\hat{\mu}},
\label{eqn:muhatRS}
\end{align}
where $\mu$ is a given parameter in the fitness function
and the right hand side depends on $\hat{\mu}$. 
The stability analysis for the RS solutions
presented by de Almeida and Thouless (AT) \cite{AT}
affords three conditions \cite{SHK2-long}:
\begin{align}
{\rm AT}_1&\equiv 1-\beta_S^2(1-2q_1+r_{11})>0\\
\nonumber
{\rm AT}_2&\equiv\Big\{1-\beta_S^2\Big(1-(\beta_J+4)q_3+(\beta_J+3)r_{33}\Big)\Big\}\\
\nonumber
&\hspace{-0.8cm}\times\Big[\beta_J+1-\beta_S^2\Big((\beta_J+1)(1-q_3)+(\beta_J-1)(q_1-r_{22})\Big)\Big]\\
&~~~+2\beta_J(\beta_J+2)\beta_S^4(q_2-r_{23})^2>0\\
{\rm AT}_{3}&\equiv1-\beta_S^2(1-2q_3+r_{33})>0,
\end{align}
where
\begin{align}
\nonumber
r_{11}&=(1-p)\langle\tanh^4(\sqrt{\hat{q}_1}u)\rangle_0+p\langle\tanh^4(\hat{\mu}+\sqrt{\hat{q}_1}u)\rangle_{\hat{\mu}}\\
\nonumber
r_{22}&=(1-p)\langle\tanh^2(\sqrt{\hat{q}_1}u)\tanh^2W(u,v)\rangle_{0}\\
\nonumber
&\hspace{0.8cm}+p\langle\tanh^2(\hat{\mu}+\sqrt{\hat{q}_1}u)\tanh^2W(u,v)\rangle_{\hat{\mu}}\\
\nonumber
r_{23}&=(1-p)\langle\tanh(\sqrt{\hat{q}_1}u)\tanh^3W(u,v)\rangle_{0}\\
\nonumber
&\hspace{1.0cm}+p\langle\tanh(\hat{\mu}+\sqrt{\hat{q}_1}u)\tanh^3
 W(u,v)\rangle_{\hat{\mu}}\\
\nonumber
r_{33}&=(1-p)\langle\tanh^4W(u,v)\rangle_0+p\langle\tanh^4W(u,v)\rangle_{\hat{\mu}}.
\end{align}

We introduce an expectation value for the target magnetization
$m_t=[\langle\sum_{i=1}^tS_i\rangle\slash t]_{\beta_J}$.
When $m_t=0$, the fitness function is also equal to 0.
Hence,
the adaptation phase is the region satisfying $m_t>0$. 
Following the replica method, the target magnetization is given by
\begin{align}
m_t=\langle\tanh W(u,v)\rangle_{\hat{\mu}},
\label{target_m}
\end{align}
which indicates that when $q_2=0$,
the target magnetization is also 0.
Thus, the parameter region with $q_2>0$ ($q_2=0$)
corresponds to the adaptation (non-adaptation) phase, respectively.

\begin{figure}
\begin{center}
\includegraphics[angle=270,width=\figwidth]{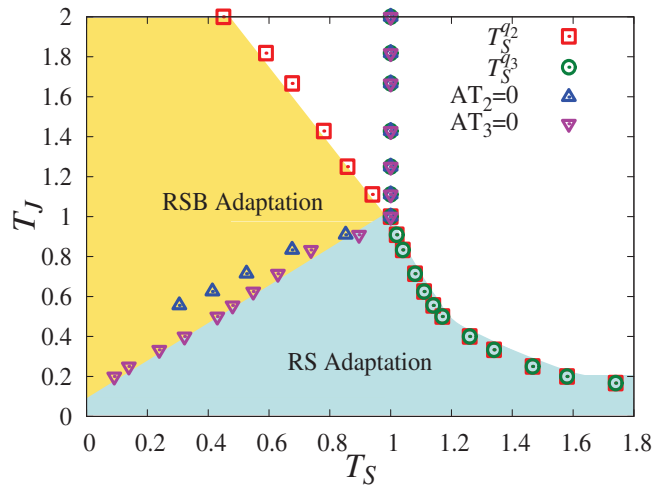}
\end{center}
\caption{(color online) Phase diagram on the $T_S-T_J$ plane at 
$p=0.2$. $\square$ and $\bigcirc$ indicate the transition temperatures
$T_S^{q_2}$ and $T_S^{q_3}$, respectively.
The adaptation phase appears in the  temperature region lower than $\square$.
$\bigtriangleup$ and $\bigtriangledown$ mark the boundary of the
RSB phase and RS phase.
}
\label{phase}
\end{figure}

The phase diagram on the $T_S-T_J$ plane at $p=0.2$ is shown in \Fref{phase}.
Here we focus our attention on the case with $\mu=1$,
and we set 
$\hat{\mu}$ to be
sufficiently large
to satisfy the self-consistent equation \Eref{eqn:muhatRS} with $\mu=1$.
%
We define the transition temperatures $T_S^{q_2}$ and $T_S^{q_3}$
such that $q_2$ and $q_3$ are positive or zero, respectively,
while $q_1$ takes a non-zero value at any finite $T_S$.
%
%
%
At $T_J>1$, the transition temperature $T_S^{q_3}$ is equal to 1 and
the temperature $T_S^{q_2}$ is smaller than $T_S^{q_3}=1$.
Adaptation 
occurs
at $T_S<T_S^{q_2}$, but
the AT stability conditions AT$_2$ and AT$_3$ are already violated at $T_S=1$.
A preliminary Monte Carlo simulation indicates that
the transition for $q_2>0$ and replica symmetry breaking (RSB) occurs at $T<T_S^{q_3}=1$ \cite{SHK2-long}.
At $T_J\leq 1$,
$T_S^{q_2}$ coincides with $T_S^{q_3}$, while RSB occurs at a lower temperature
at which $AT_3=0$. Thus
the adaptation phase $T_S\leq T_S^{q_2}$ consists of
RS and RSB phases, separated by the line AT$_3=0$.
The RS adaptation phase is thermodynamically stable at
$T_S^{\rm RSB}<T_S<T_S^{\rm RS}$,
where $T_S^{\rm RSB}$, given by  AT$_3=0$, is
the boundary between the RSB and RS phases,
and $T_S^{\rm RS}$ is the transition temperature for $q_2$ and $q_3$,
$T_S^{\rm RS}=T_S^{q_2}=T_S^{q_3}$.
As $p$ decreases, the region of the RS adaptation phase becomes narrower and eventually
the lines of AT$_2=0$, AT$_3=0$,
$T_S^{q_2}$, and $T_S^{q_3}$ merge to $T_S=1$ for any $T_J$ at $p=0$.
In this limit, the present model is identical to the SK model
whose spin-glass transition with RSB occurs at $T_S=1$
independent of $T_J$.

\begin{figure}
\begin{tabular}{cc}
\begin{minipage}{0.5\hsize}
\begin{center}
\includegraphics[width=1.8in]{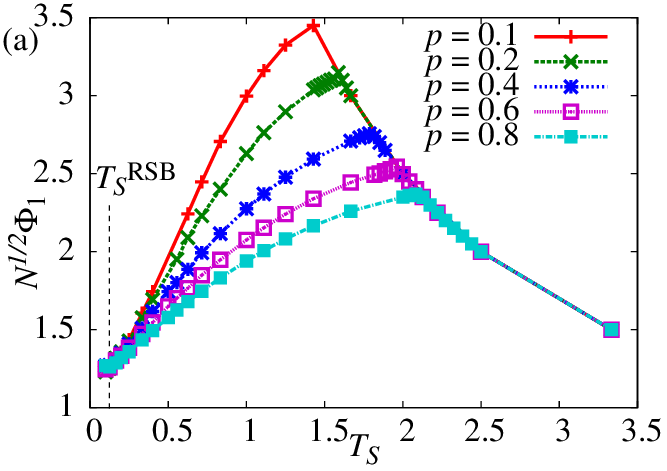}
\end{center}
\end{minipage}

\begin{minipage}{0.5\hsize}
\begin{center}
\includegraphics[width=1.8in]{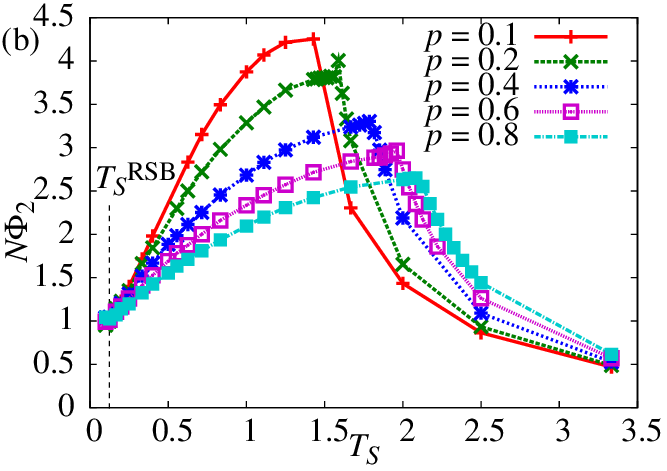}
\end{center}
\end{minipage}
\end{tabular}
\caption{(color online) $T_S$ dependence of the frustration parameters
(a) $\Phi_1$ and (b) $\Phi_2$ at $T_J=0.2$.
The vertical axis in (a) and (b) are rescaled with $\sqrt{N}$ and $N$,
respectively.
The RSB transition temperature, which weakly depends on the value of $p$ shown here,
 is indicated by the dashed lines.
}
\label{figure:frustration}
\end{figure}

To distinguish the interactions evolved in the RS phase
from those in the RSB phase,
we calculate the equilibrium frustration parameters.
Indeed, the frustration characterizes the interactions in spin glasses. 
It is defined as the product of $J_{ij}$s along a minimal loop.
When the interactions among the three spins satisfy $J_{ij}J_{jk}J_{ki}<0$,
the energy per spin cannot reach the minimum value, and such interactions are said to be frustrated \cite{Nishimori, Toulouse}.
%
In the present model,
the target spins play a distinct role
because their configuration determines the fitness function.
Hence, by distinguishing the target spins from others, we introduce the frustration parameters as
$\Phi_1\equiv\frac{1}{{\rm C}^t_2}\sum_{i<j\leq t}[J_{ij}]_{\beta_J}$ and
$\Phi_2\equiv\frac{1}{{\rm C}^t_2(N-t)}\sum_{i<j\leq t}\sum_{k=t+1}^N[J_{ik}J_{jk}]_{\beta_J}$,
where ${\rm C}^t_2$ is the number of interactions between the target spins.
When $\Phi_1=0$, the interactions between the target spins are randomly distributed;
however, when $\Phi_1>0$, ferromagnetic interactions are dominant.
The ferromagnetic interactions between the target spins energetically
favor the spin configuration with $m_t=1$.
The frustration parameter $\Phi_2$ is the average correlation of the
interactions between the target and non-target spins.
When 
the interactions 
that couple a non-target spin $S_k$ to the target spins
$S_i$ and $S_j$
satisfy the condition $J_{ik}J_{jk}>0$,
the target configuration $S_i=S_j=1$ is stable irrespective of $S_k$.
Therefore, the finite frustration parameter $\Phi_2>0$
implies that the configuration with $m_t>0$
is energetically supported by the interactions between target and
non-target interactions.
%
Under the RS ansatz, the frustration parameters are calculated 
as 
\begin{align}
\Phi_1^{\rm RS}&=\frac{\beta_S\beta_J}{\sqrt{N}}(\mu^2-m_t^2)\label{part5:average-targetJ}\\
\nonumber
\Phi_{2}^{\rm RS}&=\frac{\beta_S^2\beta_J}{N}\{(\beta_J-1)q_1^{nt}\mu^2-2\beta_Jq_2^{nt}\mu m_t\\
&~~~+(\beta_J+1)q_3^{nt}m_t^2+(\mu^2-m_t^2)\}.
\end{align}
Here the coefficients $N^{-1\slash 2}$ and $N^{-1}$ reflect the change in the order
of the interactions into $O(N^{-1\slash 2})$
through the evolution \cite{SH}.
%
%
%
%
%
%
As seen in \Fref{figure:frustration}, for any $p$,
the frustration parameters $\Phi_1$ and $\Phi_2$
increase with a decrease in $T_S$ down to $T_S =T_S^{\rm RS}$,
but with further decrease in $T_S<T_S^{\rm RS}$,  $\Phi_1$ and $\Phi_2$
decrease.
Thus, the frustration is minimal at around the
transition temperature $T_S^{\rm RS}$.
This result is consistent with the behavior of the energy \cite{SHK2-long}.
The configurations of the interactions that evolved in the intermediate temperature
range $T_S^{\rm RS}<T_S<T_S^{\rm RSB}$ have smaller frustration in the
interactions between target spins and those between target and
non-target spins.

In summary,
we employed a spin-glass model of adaptive evolution
to discuss evolutionary robustness in terms of
statistical physics.
%
Our analysis showed the existance of
two kinds of adaptation phases,
an RS adaptation phase at $T_S^{\rm RSB}<T_S<T_S^{\rm RS}$ and
an RSB adaptation phase at $T_S<T_S^{\rm RS}$.
The equilibrium properties of the
interactions were characterized by the frustration parameters, which showed that the RS adaptation phase
energetically supports the target configurations by suppressing the frustration in the evolved interactions.

Now we discuss the biological relevance of our results.
An evolved system in the RS phase is robust to noise in the dynamic processes and
to genetic change.
The relaxation dynamics of spins
progresses smoothly without becoming stuck at any metastable states.
In the RS phase, the adapted phenotype, that is, the target spin configuration, 
is a unique stable state 
that is reachable from any initial conditions
after a short time of relaxation. 
This dynamical process agrees well with  that of the funnel
landscape in protein folding \cite{Go, Onuchic},
as is also observed in evolution dynamics in biology \cite{Li,KK-Asashima}.
Next,
the self-averaging property in the RS phase guarantees
an identical equilibrium distribution of
the phenotype 
even if the genotype $\bm{J}$ is distributed around the 
evolved point.
An identical phenotype is generated irrespective of genotypic variance, which is
known as genetic canalization \cite{Waddington}. However, phenotypic robustness is lost at lower temperatures
by RSB, as represented by the appearance of a continuous overlap function.
Thus, our findings provide an evolutionary interpretation for RSB and also confirm a positive role
for thermal noise in shaping the funnel-like dynamics and robustness to mutation.

Finally, despite the use of a simple statistical-physics model of interacting spins, we expect our findings to hold true
in other problems involving evolutionary and developmental dynamics. In addition,
the proposed replica formalism could function as a theoretical
basis to understand the evolution of robustness in general.

This work was supported by Grants-in-Aid for Scientific Research
(No. 22340109 and No. 21120004) from MEXT
and for JSPS Fellows (No. 20--10778 and No. 23--4665) from JSPS.


\end{document}